\documentclass[10pt, conference]{IEEEtran}
\IEEEoverridecommandlockouts

\usepackage{cite}
\usepackage{algorithmic}
\usepackage{graphicx}
\usepackage{textcomp}
\usepackage{xcolor}

\definecolor{darkblue}{rgb}{0.,0.,0.4}
\definecolor{darkred}{rgb}{0.5,0.,0.}
\definecolor{darkpurple}{rgb}{0.5,0.,0.5}
\definecolor{ltgreen}{rgb}{0.1,.59,.43}
\definecolor{orange}{rgb}{1.0, 0.5, 0.0}
\def\BibTeX{{\rm B\kern-.05em{\sc i\kern-.025em b}\kern-.08em
		T\kern-.1667em\lower.7ex\hbox{E}\kern-.125emX}}

\usepackage[T1]{fontenc} % optional
\usepackage{amsmath}
\usepackage[cmintegrals]{newtxmath}
\usepackage{bm} % optional

\DeclareRobustCommand{\IEEEauthorrefmark}[1]{\smash{\textsuperscript{\footnotesize #1}}}
\usepackage[linktocpage=true, colorlinks=true, allcolors=blue,]{hyperref}
\usepackage{physics} % bra and ket
\usepackage{mathtools} % \vcentcolon
\newcommand{\defeq}{\vcentcolon=}
\newcommand{\eqdef}{=\vcentcolon}
\newcommand{\acr}[1]{\textsc{\lowercase{#1}}}

\DeclareMathAlphabet{\mymathbb}{U}{bbold}{m}{n}
\newcommand{\I}{{i\mkern1mu}} %{i}
\newcommand{\Id}{\mymathbb{1}}

\newcommand{\myminusspace}{\thinspace \thinspace}

\begin{document}
	\bstctlcite{IEEEexample:BSTcontrol}
	
	\title{Iterative Linear Quadratic Regulator for Quantum Optimal Control}
	
	%\thanks{
		%    This work has been submitted to the IEEE for possible publication. Copyright may be %transferred without notice, after which this version may no longer be accessible.
		%}
	
	\author{
		\IEEEauthorblockN{
			Dirk Heimann\IEEEauthorrefmark{1,}\IEEEauthorrefmark{2},
			Felix Wiebe\IEEEauthorrefmark{3},
			Tahereh Abad\IEEEauthorrefmark{4},
			Elie Mounzer\IEEEauthorrefmark{3},
			Tangyou Huang\IEEEauthorrefmark{4},\\
			Frank Kirchner \IEEEauthorrefmark{1,}\IEEEauthorrefmark{3}, and
			Shivesh Kumar\IEEEauthorrefmark{5}
		}
		\IEEEauthorblockA{\IEEEauthorrefmark{1}
			Robotics Research Group, University of Bremen
			%dirk.heimann@uni-bremen.de
		}
		\IEEEauthorblockA{\IEEEauthorrefmark{2}
			Machine Learning Research Group, Bremen University of Applied Science 
		}
		\IEEEauthorblockA{\IEEEauthorrefmark{3}
			Robotics Innovation Center, German Research Center for Artificial Intelligence, Bremen
			%elie.mounzer@dfki.de felix.wiebe@dfki.de
		}
		\IEEEauthorblockA{\IEEEauthorrefmark{4}
			Department of Microtechnology and Nanoscience, % (MC2),
			Chalmers University of Technology%, SE-41296 G\"oteborg, Sweden
		}
		\IEEEauthorblockA{\IEEEauthorrefmark{5}
			Division of Dynamics, Department of Mechanics \& Maritime Sciences,
			Chalmers University of Technology
			%Gothenburg, Sweden shivesh.kumar@chalmers.se
		}
	}
	
	\maketitle
	
	\begin{abstract}
		Quantum optimal control for gate optimization aims to provide accurate, robust, and fast pulse sequences to achieve gate fidelities on quantum systems below the error correction threshold. Many methods have been developed and successfully applied in simulation and on quantum hardware. In this paper, we establish a connection between the iterative linear quadratic regulator and quantum optimal control by adapting it to gate optimization of quantum systems. We include constraints on the controls and their derivatives to enable smoother pulses. We achieve high-fidelity simulation results for X and cross-resonance gates on one- and two-qubit fixed-frequency transmons simulated with two and three levels.
	\end{abstract}
	
	
	\section{Introduction} %incl. sota - 1.5 (qDC: 1)
	%\message{The column width is: \the\columnwidth}
	%providing solutions for a wide range of applications, including online trajectory optimization~\cite{tassa2014}	
	Quantum optimal control (QOC) aims to design and apply precisely shaped pulse sequences of external fields to achieve high-fidelity control of quantum systems as efficiently as possible~\cite{glaser2015}. While possible applications for QOC range from quantum metrology~\cite{liu2022} to quantum sensing~\cite{rembold2020}, we focus on the field of gate optimization for quantum systems~\cite{koch2022}. Quantum control algorithms build on the rich field of classical control theory, especially for trajectory optimization: Krotov's method~\cite{reich2012} is named after its classical counterpart~\cite{krotov1993}, PRONTO~\cite{shao2022} is based on the Projection Operator-based Newton method~\cite{hauser2002}, and PICO~\cite{trowbridge2023} is based on direct collocation~\cite{hargraves1987}. Iterative linear quadratic regulator (iLQR)~\cite{li2004} is a value-function-based shooting method for trajectory optimization utilizing an approximation of the Hessian to determine the control updates. In recent years, it has become a state-of-the-art method in robotics, especially in online trajectory optimization for complex robots~\cite{wensing2024} because it obtains convergence guarantees by Levenberg-Marquardt regularization, is fast because of the Hessian approximation, handles nonlinear system dynamics via line-search, and obeys the system dynamics at any iteration of the optimization. In this paper, we adapt iLQR to the task of quantum gate optimization. Our contributions include:
	%The iLQR method is a value-function-based trajectory optimization algorithm that solves optimization problems by reducing the problem to the Hamilton-Jacobi-Bellman equation (HJB). The HJB can be iteratively solved by locally approximating the discrete system dynamics linearly and the cost function quadratically.
	
	\begin{IEEEitemize}
		\item Adaptation of iLQR to QOC with smoothed controls
		\item Providing an in-depth description of the algorithm
		\item Simulation results for high-fidelity one-qubit and two-qubit gate synthesis
		for fixed-frequency transmons
		\item Comparison of our results to solutions obtained by \acr{L-BFGS-B} GRAPE
	\end{IEEEitemize}
	
	This paper is organized as follows: First, in Sec.~\ref{sec:background}, we provide a literature overview of QOC algorithms and recall the one-qubit and two-qubit fixed-frequency transmon Hamiltonian for superconducting systems. Then, in Sec.~\ref{sec:method}, we introduce iLQR with the isomorphic representation and Pad\'e approximation. In Sec.~\ref{sec:results}, we demonstrate the performance for gate optimization on the transmon Hamiltonians of increasing difficulty and conclude in Sec.~\ref{sec:conclusion}.
	
	\section{Background}\label{sec:background}
	\subsection{Quantum optimal control}\label{subsec:qoc}
	%\subsection{Related work}\label{subsec:sota}
	State-of-the-art methods for gate optimization for superconducting transmon architectures include analytic and numerical methods. Prominent analytical methods are the Derivative Removal by Adiabatic Gate (DRAG)~\cite{motzoi2009} for single-qubit operation, the cross-resonance (CR) gate~\cite{chow2011}, and its improved echo CR~\cite{sheldon2016} two-qubit operation. Both have shown remarkable results for fixed-frequency transmon architectures because of the opportunity to fine-tune the pulse shape with system feedback data~\cite{jurcevic2021}. However, numerical optimization methods promise to offer more flexibility to adapt to more complex systems, specific control problems, experimental limitations, and uncertainties~\cite{koch2022}.
	
	Many numerical open-loop control methods have been introduced in the last decades. In particular, two well-established gradient-based methods include gradient ascent pulse engineering (GRAPE)~\cite{khaneja2005} and Krotov's method~\cite{reich2012}. GRAPE updates all piecewise constant controls concurrently, i.e., based on the non-updated controls of the previous iteration step, using first-order~\cite{khaneja2005} or second-order approximations by \acr{L-BFGS-B}~\cite{machnes2011} of the loss gradients. The \acr{L-BFGS-B} GRAPE implementation is available in the \texttt{qutip-qtrl} package~\cite{johansson2012} and has been tested on IBM superconducting systems~\cite{matekole2022}. In addition, Krotov's method~\cite{reich2012}, with its quasi-Newton extension~\cite{eitan2011}, updates all controls sequentially, i.e., by including updated controls within the same iteration step. Krotov's method is available for time-continuous~\cite{tannor1992} and time-discretized control functions~\cite{goerz2019}. However, the monotonic convergence of Krotov's method is guaranteed only for the time-continuous case~\cite{reich2012}. Like Krotov's time-continuous version, PRONTO~\cite{shao2022} solves the optimal control problem by sequentially solving ordinary differential equations and is guaranteed to be monotonically convergent. In contrast to previous Krotov’s methods, PRONTO employs a Newton descent method to allow for quadratic convergence rates in proximity of local minima by incorporating system dynamics into a modified cost function using a projection operator. In addition to these shooting methods, direct collocation has recently been proposed~\cite{trowbridge2023} for quantum gate synthesis that includes both the states and controls at each discrete time step as a decision variable. Hence, in contrast to shooting methods, the number of optimization variables scales with the dimension of the Hilbert space. Additionally, the system dynamics of the trajectory are obeyed up to a certain factor by using them as a constraint during the optimization rather than a strict forward rollout. For a more comprehensive overview of recent progress in QOC, including machine learning and reinforcement learning approaches, we refer the reader to~\cite{koch2022}.
	
	The iLQR algorithm utilizes first-order derivatives of the system dynamics in a Gauss-Newton approximation step to determine the Hessian for updating piecewise constant controls. Thereby linking it closely to differential dynamic programming (DDP)~\cite{mayne1966}, which requires second-order derivatives for applying a full Newton step. However, the time advantage of the quadratic convergence often diminishes by the longer calculation time needed to determine second-order derivatives~\cite{tassa2014}. Although iLQR is concurrent in the sense that updates of piecewise constant controls are based on derivatives of the previous iteration step, a fundamental difference in comparison with GRAPE is that iLQR updates the controls based on the value function, which assigns a cost-to-go to each state and hence includes knowledge about future costs of the current control in the update.
	
	\subsection{Transmon Hamiltonian}\label{subsec:transmons}% ~1 page
	Closed quantum systems evolve according to the Schr\"odinger equation. For time-dependent unitary operators $U(t)\in \mathbb{C}^{d\times d}$ and a Hamiltonian $H$ depending on $m$ real-valued controls $u(t)\in \mathbb{R}^m$ the Schr\"odinger equation reads $\I\hbar \dot{U}(t)= H(u(t))U(t)$. We set the reduced Planck's constant $\hbar=1$ in the following. The solution for $N$ time-discretized controls $u_k\in \mathbb{R}^m$ and a constant time interval $\Delta t$ is given by
	\begin{align}\label{eq:U_final}
		U_{N}(u_{1:N}) = \prod_{k=1}^{N} \exp\left(-\I H(u_k)\Delta t\right)U_0.
	\end{align}
	The discrete system dynamics are given by
	\begin{align}\label{eq:forward_step}
		U_{k+1} = f(U_k, u_k)
		= \exp(-\I H(u_k)\Delta t)U_k.
	\end{align}            
	
	Qubits can be physically realized by fixed-frequency transmons dispersively and capacitively coupled to a quantum resonator bus~\cite{majer2007} that can be driven by microwave pulses~\cite{krantz2019}. A single transmon with transition frequency $\omega$ between eigenstates $\ket{0}$ and $\ket{1}$ and anharmonicity $\delta$ can be described by the Duffing oscillator
	\begin{align}\label{eq:duffing}
		H_\text{duffing} = \omega b^\dagger b + \frac{\delta}{2} b^\dagger b (b^\dagger b - \Id )
	\end{align}
	where $b^\dagger$ and $b$ denote the creation and annihilation operators of the quantum harmonic oscillator. The resonator is described by a harmonic oscillator Hamiltonian $\omega_r c^\dagger c$ with transition frequency $\omega_r$ for all levels, and frequencies being far detuned, i.e., $\omega_r \gg \omega$.
	
	Transmons can be coupled capacitively to the resonator
	by the Jaynes-Cummings Hamiltonian with coupling strength $g_j$
	in the dispersive regime, i.e. $g_j \ll \abs{\Delta_{jr}}$
	where $\Delta_{jr} = \omega_j - \omega_r$ is the detuning.
	For two transmons coupled to the resonator, the Hamiltonian reads
	\begin{align}
		%\begin{split}
		H_\textbf{sys} &= \omega_r c^\dagger c + \sum_{j=1}^2H_\text{duffing, j}
		+ \sum_{j=1}^2 g_j(b_j^\dagger c + b_jc^\dagger)
		%H_\textbf{sys} &= \omega_r c^\dagger c + \sum_{j=1}^2 \left(
		%\omega_j b_j^\dagger b_j + \frac{\delta_j}{2} b_j^\dagger b_j (b_j^\dagger %b_j - \Id ) \right) \\
		%& + \sum_{j=1}^2 g_j(b_j^\dagger c + b_jc^\dagger).
		%\end{split}
	\end{align} %The first seven terms
	The first resonator term and the six Duffing terms of the system's Hamiltonian are diagonal. Only the last two interaction terms are non-diagonal. Therefore, in the dispersive regime, the following effective two-transmon Hamiltonian can be derived via a Schrieffer-Wolf transformation and projection on the zero-excitation subspace of the bus~\cite{magesan2020}
	\begin{align}
		%\begin{split}
		H^\text{eff}_\text{sys} &= \sum_{j=1}^2 \left(
		\tilde{\omega}_j b_j^\dagger b_j + \frac{\delta_j}{2} b_j^\dagger b_j (b_j^\dagger b_j - \Id ) \right) %\nonumber \\
		+ J_{12}\left( b_1^\dagger b_2 + b_1 b_2^\dagger\right)
		%\end{split}
	\end{align}
	with dressed qubit frequencies $\tilde{\omega}_j = \omega_j + \frac{g_j^2}{\Delta_{jr}}$ and
	the effective coupling to first-order $J_{12} = \frac{g_1g_2}{2\Delta_{1r}\Delta_{2r}}(\Delta_{1r} + \Delta_{2r})$.
	
	General microwave drive channels acting on the $j$th transmon with 
	in-phase quadrature $u_j^X(t)$, off-phase quadrature $u_j^Y(t)$, drive frequency $\omega_d$, and phase $\phi$
	are modeled by the drive Hamiltonian
	\begin{align}
		H_\text{d}(t) &=
		r_j %\nonumber \\
		\left(u^X_j(t) \cos{(\omega_d t + \phi)} + u^Y_j(t)\sin{(\omega_d t + \phi)}\right)
		(b^\dagger_j + b_j)
	\end{align}
	where we include the Rabi strength factor $r_j$, which acts as a damping factor and is a property of the quantum system.
	
	In the single-transmon case, the transmon is driven with two quadratures %$u^X(t)$ and $u_k^Y$
	%$H_\text{drive, 1-t} =\left(\Omega^X(t) \cos{(\omega t)} + \Omega^Y(t)\sin{(\omega t)}\right)(b^\dagger + b)$
	in its dressed transition frequency $\omega_d = \tilde{\omega}$ and the total system is described by $H_\text{duffing} + H_d(t)$. The transformation $ R = \exp(-\I (\omega_d t + \phi) b^\dagger b) $ shifts to the rotating frame,e which leads to the equation of one transmon driven by a microwave pulse with the rotating frame approximation (RWA)
	\begin{align}\label{eq:1-t}
		H^\text{RF}_\text{1-t}(t) &= \frac{\delta_1}{2} b_1^\dagger b_1 (b_1^\dagger b_1 - \Id ) \nonumber \\ 
		&+ r_1\frac{u^X(t)}{2} (b_1^\dagger + b_1) + r_1\frac{u^Y(t)}{2} \I (b_1^\dagger - b_1).
	\end{align}
	In the two-transmon case, a second drive can be added that acts on the second transmon $(b^\dagger_2 + b_2)$ at the frequency of the first $\omega_d = \tilde{\omega}_1$. Driving one transmon (the control qubit) with the frequency of the other (the target qubit) is referred to as cross resonance pulse~\cite{chow2011}. Both drives have the same frequency, i.e., $\tilde{\omega}_1$, but act on different transmons. This enables a rotating frame transformation by $ R = \exp(-\I (\tilde{\omega}_1t + \phi) \sum_{j=1}^2 b^\dagger_j b_j )$. With RWA and $\Delta_{21} = \tilde{\omega}_2 - \tilde{\omega}_1$, the two-transmon Hamiltonian is given by %Utilizing the
	\begin{align}\label{eq:2-t}
		H^\text{RF}_\text{2-t}(t) &= \Delta_{21} b_2^\dagger b_2
		+ \sum_{j=1}^2 \frac{\delta_j}{2} b_j^\dagger b_j (b_j^\dagger b_j - \Id )
		\nonumber %\\ &
		+ J_{12}\left( b_1^\dagger b_2 + b_1 b_2^\dagger\right)
		\nonumber \\
		&+ r_1\frac{u^{X_1}(t)}{2} (b_1^\dagger + b_1) + r_1\frac{u^{Y_1}(t)}{2} \I (b_1^\dagger - b_1) \nonumber \\
		&+ r_1\frac{u^{X_2}(t)}{2} (b_2^\dagger + b_2) + r_1\frac{u^{Y_2}(t)}{2} \I (b_2^\dagger - b_2).
	\end{align}
	%where $\Delta_{21}$ is the difference between the dressed frequencies $\tilde{\omega}_2 - \tilde{\omega}_1$.
	
	\subsection{Hardware parameters}
	On quantum systems, the arbitrary waveform generator (AWG) creates the amplitudes for the pulses as piecewise constant functions $u^{X_j}_k$ and $u^{Y_j}_k$ which are
	applied for the duration $\Delta t$ for each timestep $k\in \{1, \ldots, N\}$. We utilize the system parameters of the IBM Eagle processor \texttt{ibm\_brisbane} for qubit 0 and 1, which can be obtained via the open source library \texttt{Qiskit Pulse}~\cite{alexander2020}. Table~\ref{tab:parameters} lists the rounded parameters in angular frequencies.

	%IBM Eagle processors have a native gate set
	%In IBM Eagle processors, the current gate times are
	%$80\:$ns for $\sqrt{X}$ and $X$, and $660\:$ns for the ECR gate
	%which consists of $300\:$ns $\operatorname{CR}(\pi/4)$ pulses, 
	%a $60\:$ns single-qubit drag pulse, and 
	%$300\:$ns $\operatorname{CR}(-\pi/4)$ pulses across different drive channels.
	A common native gate set for fixed-frequency transmon architectures consists of single-qubit operations $\sqrt{X}$, $X$, and $R_Z(\theta)$ and an entangling two-qubit operation like cross resonance, echoed cross resonance, or CNOT gate. The $R_Z(\theta)$ gate can be implemented virtually by changing the phase of the AWG~\cite{mcKay2017}. The single qubit gates $\sqrt{X}$, $X$ and the two-qubit operation are implemented by AWG pulse envelopes. We focus on the X gate for the one-qubit operation because the method works analogously for the $\sqrt{X}$ gate. For the two-qubit operation, we consider the cross-resonance gate $R_{XZ}(\pi/2) = \exp(-\I \pi XZ /4)$, which is equivalent to the CNOT gate up to a global phase by applying it together with one-qubit gates $CX_{q_2q_1} = \Id \otimes R_Z(-\pi/2) \cdot R_{XZ}(\pi/2) \cdot R_X(-\pi/2)\otimes \Id$. Note that, in our drive channel setup, the second qubit is the control and the first qubit is the target.
	
	\begin{table}[b]
		%\renewcommand{\arraystretch}{1.0}
		%\caption{Parameters for the transmon Hamiltonian in angular frequencies $2\pi \text{GHz}$.}
		\caption{Fixed-frequency transmon parameters}
		\label{tab:parameters}
		\centering
		\begin{tabular}{lcl}
			\hline %\toprule
			%ibm_brisbane_2025-02-11_13-04
			qubit 1 dressed frequency & $\tilde{\omega}_1/(2\pi)$ & \myminusspace 4.7219 {\upshape GHz}\\
			%721908316964448
			qubit 2 dressed frequency & $\tilde{\omega}_2/(2\pi)$ & \myminusspace 4.8151 {\upshape GHz}\\
			%815135900326559
			qubit 1 anharmonicity     & $\delta_1/(2\pi)$         &              -0.3120 {\upshape GHz}\\
			%31197865973435573
			qubit 2 anharmonicity     & $\delta_2/(2\pi)$         &              -0.3097 {\upshape GHz}\\
			%30974403197770844
			effective coupling        & $J_{12}/(2\pi)$           & \myminusspace 0.0020 {\upshape GHz}\\
			%0020343864324431304
			qubit 1 rabi strength     & $r_1/(2\pi)$              & \myminusspace 0.0921 {\upshape GHz}\\
			%09212484141794322
			qubit 2 rabi strength     & $r_2/(2\pi)$              & \myminusspace 0.0974 {\upshape GHz}\\
			%09744987921644638
			minimal time interval     & $\Delta t$         & \myminusspace 0.5 {\upshape ns}\\
			\hline %\bottomrule
		\end{tabular}
	\end{table}
	
	\section{Method}\label{sec:method} % to avoid calculating with complex numbers
	This section outlines the isomorphic representation, the Pad\'e approximation for approximating the matrix exponential, the iLQR method, and its adaptations to obtain smoother control shapes.
	
	\subsection{Isomorphic representation} %\cite{palao2003, shao2022,trowbridge2023}
	We follow~\cite{trowbridge2023} and vectorize unitary matrices by an isomorphic representation of complex vector spaces. A complex-valued vector $\ket{\Psi} \in \mathbb{C}^d$ and matrix $U \in \mathbb{C}^{d \times d}$ can be rewritten as the following real-valued vector and matrix
	\begin{align}
		\ket{\tilde{\Psi}} =
		\begin{pmatrix}
			\Re \Psi \\
			\Im \Psi
		\end{pmatrix} \quad \text{and} \quad
		\tilde{U} =
		\begin{pmatrix}
			\Re U & - \Im U \\
			\Im U & \Re U
		\end{pmatrix}.
	\end{align}
	The state space of complex-valued matrices can, therefore, be expressed by a $2d^2$ real-valued vector
	$x_k = \begin{pmatrix}
		\Re(U_{00}), \ldots, \Re(U_{dd}), %\Re(U_{01}), 
		\Im(U_{00}), \ldots, \Im(U_{dd}) %\Im(U_{01}), 
	\end{pmatrix}^T$
	where $x_k$ denotes the matrix values at timestep $k$. The discretized forward step of Eq.~\eqref{eq:forward_step} based on controls $u_k$ can also be expressed in the isomorphic representation
	\begin{align}\label{eq:iso_forward_step}
		\tilde{U}_{k+1} = f(\tilde{U}_k, u_k)
		= \exp\left(-\I \tilde{H}(u_k)\Delta t\right)\tilde{U}_k.
	\end{align}
	
	\subsection{Pad\'e approximation}
	
	Calculating the matrix exponential in Eq.~\eqref{eq:forward_step} is essential for the forward propagation. While (quasi-)analytic expressions are available for Lie groups $SU(2)$, $SU(3)$, and $SU(4)$~\cite{kaiser2022}, we decide to use the Pad\'e approximation for matrix exponentials as described in~\cite{trowbridge2023} because it can be readily used for higher levels. The Pad\'e approximation is achieved by
	\begin{align}\label{eq:pade}
		\exp(G) \approx B^{-1}(G) F(G)
	\end{align}
	where $B(G)$ and $F(G)$ are power series that can be chosen to a certain degree. For example, the eighth-order power series is given by
	\begin{subequations}
		\begin{align}	
			B(G, \Delta t) &= \Id - \frac{\Delta_t}{2} G + \frac{3\Delta_t^2}{28}G^2
			- \frac{\Delta_t^3}{84} G^3 + \frac{\Delta_t^4}{1680} G^4 \\
			F(G, \Delta t) &= \Id + \frac{\Delta_t}{2} G + \frac{3\Delta_t^2}{28}G^2
			+ \frac{\Delta_t^3}{84} G^3 + \frac{\Delta_t^4}{1680} G^4.
		\end{align}
	\end{subequations}
	
	\subsection{Iterative Linear Quadratic Regulator}\label{subsec:ilqr} % ~1 page
	With the isomorphic representation of the forward step in Eq.~\eqref{eq:iso_forward_step}, the Pad\'e approximation in Eq.~\eqref{eq:pade}, and the vectorized form $x_k$ of the unitary $U_k$, the QOC problem can be formulated as an iLQR problem. As a trajectory optimization method, iLQR encodes the search for a control trajectory $u_{1:N-1} = (u_1,\ldots, u_{N-1})$ with a consistent state trajectory $x_{1:N} = (x_1,\ldots, x_{N})$ over a horizon $N$ which fulfills constraints in an optimization problem of the shape %certain
	\begin{subequations}
		\begin{align}
			\min_{u_{1:N-1}} &J(x_{1:N}, u_{1:N-1}) = l_f(x_N) +
			\sum_{k=1}^{N-1} l (x_k, u_k)
			\label{eq:trajopt}\\
			\text{subject to:}& \quad x_{k+1} = f(x_k, u_k) \quad \forall k \in \{1,
			..., N-1\} \\
			& \quad x_1 = \tilde{x}_1
		\end{align}
	\end{subequations}
	with the cost functional $J$, which usually can be split into a final cost function $l_n(x_N)$ and a running cost function $l(x_k, u_k)$. The constraints enforce the discrete system dynamics between the knot points and fix the initial state.
	
	%Iterative linear quadratic regulator (iLQR)~\cite{li2004} is a value function
	%based trajectory optimization algorithm that solves the discrete optimization
	%problem \ref{eq:trajopt} by reducing the problem to the Hamilton-Jacobi-Bellman
	%equation (HJB).
	%The HJB can be iteratively solved by locally approximating the discrete
	%system dynamics linearly and the cost function quadratically. 

	% Let $J(X_i, U_i)$ describe the discrete cost function for 
	% a horizon of $N-i$ time steps
	% \begin{align}\label{eq:ilqr_cost_function}
		% 	J(X_i, U_i) &= l_f(x_N) + \sum_{t=i}^{N-1} l (x_t, u_t) 
		% \end{align}
	% with final costs $l_f(x_N)$, running costs $l(x_i, u_i)$,
	% trajectories $X_i = (x_i, \ldots, x_N)$, and
	% $U_i = (u_i, \ldots, u_N)$ from step $i$ to the end $N$. \elie{?}
	The value function $V$ is defined as $V_k \defeq V(x_k) = \min_{u_{k:N-1}}
	J(x_{k:N},u_{k:N-1})$. Bellman's optimality principle states that for
	optimal solutions every individual step has to be optimal, hence yields the
	recursive relation 
	\begin{align}
		V_k(x_k) &=	\min_{u_k}\left[ l(x_k, u_k) + V_{k+1}(f(x_k, u_k))\right] \nonumber \\
		&\defeq \min_{u_k} Q_k(x_k, u_k)
	\end{align}
	with $V_N = l_f(x_N)$. The argument of the minimization is the
	action-value-function $Q(x_k, u_k)$.
	% \begin{align}
		% Q(x,u) = l(x_i,u_i) + V(f(X_{i}, u_i))
		% \end{align}
	For small variations, $Q_k(x_k + \delta x_k, u_k + \delta u_k)$ can be
	approximated with a second-order Taylor expansion, which requires the following
	terms
	%%% same as \begin{subequations} \begin{align}
			\begin{IEEEeqnarray}{rll}\label{eq:ilqr}
				Q_{x,k} \; &=  \; l_{x,k}     &+ \; V_{x,k+1} f_{x,k}
				\IEEEyesnumber\IEEEyessubnumber* \\
				Q_{u,k} \; &=  \; l_{u,k}     &+ \;V_{x,k+1} f_{u,k} \\
				Q_{xx,k} \; &=  \; l_{xx,k} &+ \; f_{x,k}^T V_{xx,k+1} f_{x,k} + V_{x,k+1}f_{xx,k} \\
				Q_{uu,k} \; &=  \; l_{uu,k} &+ \; f_{u,k}^T V_{xx,k+1} f_{u,k} + V_{x,k+1}f_{uu,k} \\
				Q_{ux,k} \; &=  \; l_{ux,k} &+ \; f_{u,k}^T V_{xx,k+1} f_{x,k} + V_{x,k+1}f_{ux,k} \\
				Q_{xu,k} \; &=  \; Q_{ux,k}^T
			\end{IEEEeqnarray}
			where $Q_{x,k} = \partial_x Q \rvert_{x_k, u_k}$,
			$Q_{xx,k}= \partial^2_x Q \rvert_{x_k, u_k}$, and accordingly for all other combinations.
			DDP utilizes the full second-order approximation, while iLQR ignores the
			second-order derivatives of the dynamics function.
			%$V_{x, i+1} = \partial_x V\rvert_{x_{i+1}}$, and
			%$f_{x,i} = \partial_x f\rvert_{x_i, u_i}$.
			With this approximation, the Q-function can be minimized with respect to $\delta u$ yielding the locally optimal update to the controls
			\begin{align}
				\delta u^\star_k
				&= \operatorname{argmin}_{\delta u_k}{Q(x_k + \delta x_k, u_k + \delta u_k)} \nonumber\\
				&= - Q_{uu,k}^{-1}Q_{u,k} - Q_{uu,k}^{-1}Q_{ux,k}\delta x_k\nonumber\\
				&\eqdef \kappa_k + K_k\delta x_k
			\end{align}
			The value function can be updated by
			$V_{x,k} = Q_{x,k} - Q_{u,k} Q_{uu,k}^{-1}Q_{ux,k}$ and
			$V_{xx,k} = Q_{xx,k} - Q_{xu,k} Q_{uu,k}^{-1}Q_{ux,k}$.
			% \begin{align}
				%     V_{x,k} &= Q_{x,k} - Q_{u,k} Q_{uu,k}^{-1}Q_{ux,k} \\
				%     V_{xx,k} &= Q_{xx,k} - Q_{xu,k} Q_{uu,k}^{-1}Q_{ux,k}.
				% \end{align}
			Using a backwards recursion for $V_k$ and $V_{k+1}$ starting from the final condition $V_N = l_f(x_N)$ one can recursively compute updates to the value function and the feedback gains $\kappa_k$ and $K_k$. This procedure is called the backward pass. The invertibility of $Q_{uu}$ is ensured by including a Levenberg-Marquardt regularization scheme~\cite{mastalli2022}. During the forward pass, the feedback gains are applied and a new control and state trajectory is calculated starting from the fixed start state $x_1^\text{new} = x^1$
			\begin{align}
				u^\text{new}_k &= u_k + \alpha \kappa_k + K_k(x^\text{new}_k - x_k) \\
				%u^\text{new}_k &= u_k -Q_{uu, i}^{-1}Q_{u, i} - Q_{uu,k}^{-1}Q_{ux,k}(x^\text{new}_k - x_k) \\
				x^\text{new}_{k+1} &= f(x^\text{new}_k, u^\text{new}_k)
			\end{align}
			A line-search with the parameter $\alpha$ and a Goldstein acceptance criteria~\cite{mastalli2022} prevent steps from going too far away from the reference point around which the system dynamics are linearly approximated. Backward pass and forward pass are alternated until convergence. While the derivatives in the RHS of Eqs.~\eqref{eq:ilqr} must be calculated backward for each timestep $k$, these calculations can be performed in parallel.
			
			\subsection{Smooth controls}
			The original iLQR implementation, as presented in the previous section, does not include explicit constraints and, hence, optimizes over the entire, unconstrained control space. Because of limitations of the AWG abrupt changes in the controls should be avoided and starting and ending at zero is beneficial. To retrieve results with these properties, we choose the derivatives of the pulse envelopes as control variables and extend the state space with the actual pulse values. The derivative $\dot{u}_k$ changes the actual controls via $u_{k+1} = u_k + \dot{u}_k\Delta t$ for all $k$. With these modifications, we ensure that pulses start at zero by setting the initial condition and enable the penalization of the final pulse value, all intermediate pulse values, and the rate of change for the pulse in the cost function. Besides that, we do not impose further constraints on the controls. The total problem formulation in the structure of Eq.~\eqref{eq:trajopt} can be summarized as
			\begin{subequations}
				\begin{align}	
					\min_{\dot{u}_1, \ldots, \dot{u}_{N-1}} \quad & \mathcal{L}_N(x_N, u_N)
					+ \sum_{k=1}^{N-1}\mathcal{L}_k(x_k, u_k, \dot{u}_{k}) \\
					\text{subject to:}& \quad \tilde{U}_{k+1} = e^{-i \tilde{H}(u_k) \Delta t}\tilde{U}_k \quad \forall k \\
					&\quad u_{k+1} = u_k + \dot{u}_k \Delta t \quad \forall k \\
					&\quad \tilde{U}_1 = \Id \quad \text{and} \quad u_1 = 0.
				\end{align}
				
			\end{subequations}
			We use the quadratic costs
			\begin{align}
				\mathcal{L}_k(x_k, u_k, \dot{u}_k) &= \dot{u}_k^T R_d \dot{u}_k + u_k^T R_c u_k\\
				\mathcal{L}_N(x_N, u_N) &= (x_N - x_g)^T Q_f (x_N - x_g) +  u_N^T R_f u_N
			\end{align}
			with the cost matrices $Q_f \in \mathbb{R}^{2d^2\times 2d^2}$ and $R_d, R_c,
			R_f \in \mathbb{R}^{m\times m}$. 
			$R_d$ regularizes the rate of change for the pulses, $R_c$ the pulse
			amplitudes and $R_f$ the final control values $u_N$.
			$Q_f$ penalizes differences between the final unitary $\tilde{U}_N =
			\prod_{k=1}^{N-1} \text{exp}(-i \tilde{H}(u_k)\Delta t) \tilde{U}_1$ and the
			goal matrix $\tilde{U}_g$ computed in their vectorized form $x_N - x_g$.
			% For the running cost we use
			% \begin{align}
				% 	\mathcal{L}_k(\boldsymbol{U}_k, \boldsymbol{\Omega}_{k}, \dot{\boldsymbol{\Omega}}_{k}) &=
				% 	\boldsymbol{\Omega}_k^T
				%     R_c
				%     \boldsymbol{\Omega}_k +
				% 	\dot{\boldsymbol{\Omega}}_k^T
				%     R_d
				%     \dot{\boldsymbol{\Omega}}_k
				% \end{align}
			% with the cost matrices $R_c\in \mathbb{R}^{m\times m}$ to favor low control values for all intermediate timesteps $t_k$,
			% and $R_d\in \mathbb{R}^{m\times m}$ to favor low derivative values
			% of the control.

			\section{Results}\label{sec:results} % ~1 page
			%$80\:$ns for $\sqrt{X}$ and $X$, and $660\:$ns
			%leading to pulse durations of $40\:$ns 
			%which were both determined experimentally by hyperparameter tuning.
			This section presents the simulation results of four gate optimization examples based on superconducting transmons restricted to two and three levels. We utilize the system parameters of the \texttt{IBM\_brisbane} Eagle processor as summarized in table~\ref{tab:parameters} and keep the minimal time interval $\Delta t=0.5\:$ns constant. In addition, we start our numerical experiments by selecting the number of timesteps slightly shorter than the gate times implemented on the \texttt{IBM\_brisbane} system for their one- and two-qubit gates. We choose $N=80$ time steps for our 1-qubit experiments and $N=480$ for our 2-qubit experiments. We initialize all pulses with random values between $[-0.01, \ldots 0.01]$ and performed hyperparameter tuning of the cost matrices. As a first step, we manually find four diagonal values for the cost matrices $Q_f, R_d, R_c$, and $R_f$, which lead to good fidelities and relatively smooth pulses. For the one-qubit and three-level model, as well as for both two-qubit models, we perform a grid search by multiplying the four diagonal values by $\{1/10, 1/2, 1, 5, 10\}$ to search through $625$ different parameter combinations. For the two-qubit model with three level, we select the best run of the previous gridsearch, fix $R_c$, and define a finer grid $\{1/10, 1/4, 1/2, 3/4, 1, 2.5, 5, 7.5, 10\}$ for the remaining diagonal values of the three cost matrices. The remaining section presents the best results regarding gate fidelity and the smoothness of the envelopes.
			
			\subsection{One qubit with two levels}
			%\begin{align}
			%	U_\text{g}
			%	&= \prod_{k=0}^{N-1}\exp(-i\pi r_1u^X_k\Delta t \sigma_X) \\
			%	&= \exp(\I f \sigma_X) = \cos{(f)}\Id + \I \sin{(f)}\sigma_X \\
			%	&\overset{!}{=} \I \sigma_X \Rightarrow
			%    f = -\pi r_1 \sum_ku^X_k\Delta t \overset{!}{=} \frac{\pi}{2} \\
			%    &\Rightarrow \sum_ku^X_k\Delta t = -\frac{1}{2r_1}
			%\end{align}
			As a first example, we consider the two-level time-discretized Hamiltonian of Eq.~\eqref{eq:1-t} with piecewise constant controls. We choose $U_\text{g} = \I \sigma_X \in SU(2)$ which equals the $X$-gate up to a global phase. This task is analytically solvable because only the $u^X \sigma_X$ operation is required~\cite{motzoi2009}: if $U_\text{g} = \I \sigma_X \overset{!}{=} U_\text{N} = \prod_{k=0}^{N-1}\exp(-i\frac{r_1}{2}u^X_k\Delta t \sigma_X)$ than $\sum_ku^X_k\Delta t = -\frac{\pi}{r_1}$ because of Euler's formula for $SU(2)$. Without including the derivatives of the controls, iLQR reproduces the analytic \textit{bang-bang} solution with constant controls $u^X_k = -0.135722$ for all $N=80$ time steps $k$ leading to a simulation rollout infidelity of $1.3\cdot 10^{-13}$. The sum over all controls matches the analytic factor up to $3\cdot 10^{-7}$. However, rapid changes of the AWG's amplitude, as they can, for example, appear if the controls start and end with values different from zero, should be avoided in practice. That is why smoother controls are preferred in practice.
			
\begin{figure}[t]
				\centering
				\includegraphics[]{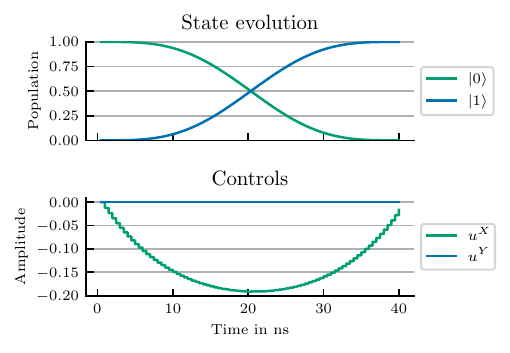}
				\caption{Simulation results for one transmon qubit with two levels with gate infidelity of $4.0\cdot 10^{-9}$. Population probability obtained by state evolution of the basis state $\ket{0}$ (\textbf{top}) under the optimal control sequence for the $X$-gate (\textbf{bottom}).}
				\label{fig:1t_2l}
			\end{figure}

			Fig.~\ref{fig:1t_2l} shows the resulting controls and state propagation for basis states $\ket{0}$ and $\ket{1}$ obtained in simulation by iLQR including the derivatives of the controls for the smoothening effect. The simulation rollout infidelity is $4\cdot 10^{-9}$ and the sum of the controls matches the factor up to $5.7\cdot 10^{-5}$.
			
			Similar results can be obtained with \acr{L-BFGS-B} GRAPE. We modified the code slightly to allow different initial pulses for the controls. Starting with a Gaussian initial pulse type for $u^X$ and zeros for $u^Y$ leads to an almost ideal Gauss-like solution with the sum being equal to the factor up to $3.9\cdot 10^{-7}$ and a simulation rollout infidelity of $1.9\cdot 10^{-13}$. The Gaussian form, compared to the solution obtained by iLQR, has the advantage of smoother start and end regions.
			
			\subsection{One qubit with three levels}
			
			%For three levels, the creation and annihilation operators are given by
			%\begin{align}
			%	a^\dagger = 
			%	\begin{pmatrix}
				%		0        & 0        & 0 \\
				%		\sqrt{1} & 0        & 0 \\
				%		0        & \sqrt{2} & 0 
				%	\end{pmatrix} \qquad
			%	a = 
			%	\begin{pmatrix}
				%		0 & \sqrt{1} & 0        \\
				%		0 & 0        & \sqrt{2} \\
				%		0 & 0        & 0 
				%	\end{pmatrix}.
			%\end{align}
			In the next example, we enable the possibility of leakage into a higher level by utilizing the time-discretized Hamiltonian of Eq.~\eqref{eq:1-t} with piecewise constant controls and three-dimensional creation and annihilation operators. We choose
			\begin{align}
				U_\text{g} = \I \left(\ket{1}\bra{0} + \ket{0}\bra{1}\right) + \ket{2}\bra{2} \in SU(3)
				%	\begin{pmatrix}
					%		0 & 1 & 0 \\
					%		1 & 0 & 0 \\
					%		0 & 0 & \I 
					%	\end{pmatrix} \in SU(3)
			\end{align}
			which acts on the basis states $\ket{0}$ and $\ket{1}$ as an ordinary $X$-gate up to a global phase. We use the derivative control to ensure a smoother rise and decline in the amplitudes.
			\begin{figure}[t]
				\centering
				\includegraphics[]{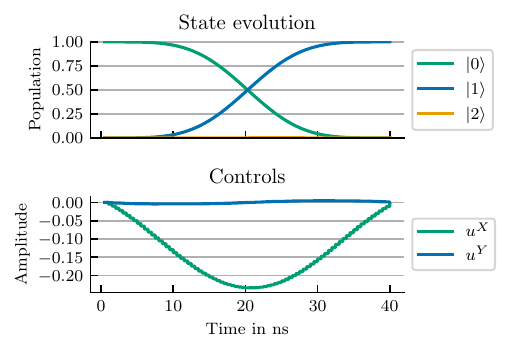}
				\caption{Simulation results for one transmon qubit with three levels
					with gate infidelity of $2.1\cdot 10^{-7}$.
					Population probability obtained by state evolution of basis state $\ket{0}$ (\textbf{top})
					under the optimal control sequence for the $X$-gate (\textbf{bottom}).}
				\label{fig:1t_3l}
			\end{figure}
			Fig.~\ref{fig:1t_3l} shows the resulting controls and state propagation for basis states |0⟩ obtained in simulation by iLQR. The infidelity is $2.1\cdot 10^{-7}$ and the sum over all controls $\sum_ku^X_k\Delta t$ equals the factor up to $3.8\cdot 10^{-4}$. The state propagation shows almost no leakage into the higher excitation state $\ket{2}$. The off-phase quadrature amplitude $u^Y$ is roughly proportional to the derivative of the $u^X$ amplitude, i.e. $u^Y = -\delta_1\dot{u}^X$, which is, up to the factor,  in agreement with the DRAG procedure~\cite{motzoi2009}. In our iLQR setup, we force both in-phase and off-phase quadratures to go towards zero at the start and end. For many hyperparameters and system values, iLQR provides $u^X$ solutions that are more parabola- than Gaussian-shaped. For parabola solutions for $u^X$, the derivatives at the start and end are highest, which means that $u^Y$, especially at the beginning and end, is not proportional to the derivative.
			
			As a comparison, when initializing \acr{L-BFGS-B} GRAPE with a Gaussian pulse shape for $u^X$ and $u^Y = -\delta_1\dot{u}^X$, we obtain solutions that mainly stay in their initial shapes with a gate infidelity of $1.6\cdot 10^{-10}$.
			%and the sum over $u^X$ being equal to the factor up to $4.6\cdot 10^{-4}$.            
			
			\subsection{Two qubits with two levels}
\begin{figure}[t]
				\centering
				\includegraphics[]{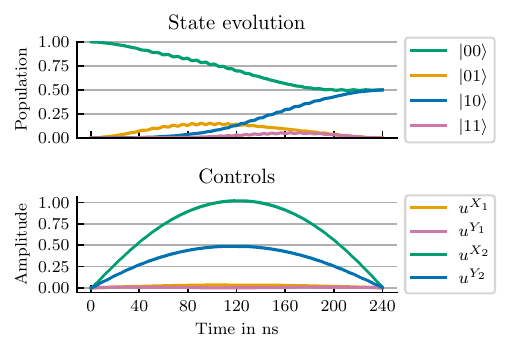}
				\caption{Simulation results for two transmon qubits with two levels
					with gate infidelity of $1.1\cdot 10^{-8}$.
					Population probability obtained by state evolution of basis state $\ket{00}$ (\textbf{top})
					under the optimal control sequence for the $240\:$ns $XZ$-gate (\textbf{bottom}).}
				\label{fig:2t_2l}
			\end{figure}

			For the two-qubit gate experiments, we start by restricting the Hamiltonian in Eq.~\eqref{eq:2-t} to two levels and choose to optimize for the cross-resonance gate
			\begin{align}
				U_\text{g} = \exp(-\I\frac{\pi}{4}\sigma_X\otimes\sigma_Z)
				\in SU(4)
				%= \frac{1}{\sqrt{2}}
				%\begin{pmatrix}
				%	1   & 0  & -\I & 0  \\
				%	0   & 1  & 0   & \I \\
				%	-\I & 0  & 1   & 0  \\
				%	0   & \I & 0   & 1 
				%\end{pmatrix}% \in SU(4)
			\end{align}
			where the first qubit is the target qubit and the second is the control qubit. Fig.~\ref{fig:2t_2l} shows the resulting controls and state propagation for basis state $\ket{00}$ obtained in simulation by iLQR. The infidelity is $1.1\cdot 10^{-8}$ with a gate duration of $240\:$ns. Four configurations in the grid search lead to lower infidelities than $10^{-7}$. They have slightly less smooth controls with varying amplitudes for all four controls such that the state's population probability evolves more periodically until it reaches its goal. %Generallyl, using the control's derivatives ensured a smoother rise and decline in the amplitudes. 
			
			\subsection{Two qubits with three levels}
\begin{figure}[t]
				\centering
				\includegraphics[]{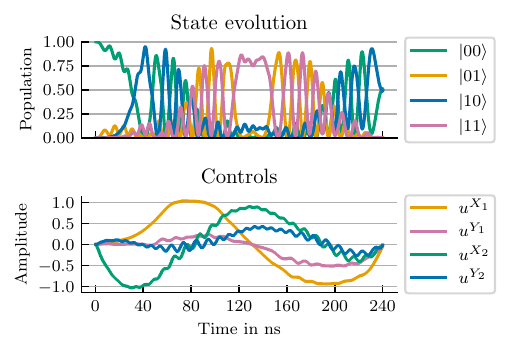}
				\caption{Simulation results for two transmon qubits with three levels
					with gate infidelity of $5.9\cdot 10^{-5}$.
					Population probability obtained by state evolution of basis state $\ket{00}$ (\textbf{top})
					under the optimal control sequence for the $240\:$ns $XZ$-gate (\textbf{bottom}).
					In the top plot, only the 2-level states are depicted because
					the contribution of the third level was much smaller with an average probability of $5.8\cdot 10^{-3}$ and a max value of $5.8\cdot 10^{-2}$ of all five higher-level basis states.}
				\label{fig:2t_3l}
			\end{figure}

			As our last example, we consider the two-transmon time-discretized Hamiltonian of Eq.~\eqref{eq:2-t} restricted to three levels. We choose to optimize the following extended CR gate %cross-resonance gate
			\begin{align}
				%\begin{split}
				U_\text{g}
				&= \sum_{j\in\{0,1,3,4\}} \frac{1}{\sqrt{2}} \ket{j}\bra{j} 
				+ \sum_{j\in \{2,5,6,7,8\}} \ket{j}\bra{j} \nonumber \\
				&+ \frac{-\I}{\sqrt{2}} (\ket{3}\bra{0} + \ket{0}\bra{3})
				+ \frac{\I}{\sqrt{2}} (\ket{4}\bra{1} + \ket{1}\bra{4})
				%\end{split}
			\end{align}
			%\begin{align}
			%	U_\text{g} = \frac{1}{\sqrt{2}}
			%	\begin{pmatrix}
				%		1   & 0  & 0        & -\I & 0  & 0        & 0        & 0       \\
				%		0   & 1  & 0        & 0   & \I & 0        & 0        & 0       \\
				%        0   & 0  & \sqrt{2} & 0   & 0  & 0        & 0        & 0       \\
				%		-\I & 0  & 0        & 1   & 0  & 0        & 0        & 0       \\
				%		0   & \I & 0        & 0   & 1  & 0        & 0        & 0       \\
				%        0   & 0  & 0        & 0   & 0  & \sqrt{2} & 0        & 0       \\
				%        0   & 0  & 0        & 0   & 0  & 0        & \sqrt{2} & 0       \\
				%        0   & 0  & 0        & 0   & 0  & 0        & 0        & \sqrt{2}\\
				%	\end{pmatrix}% \in SU(4)
			%\end{align}
			which is an element of $SU(9)$. Fig.~\ref{fig:2t_3l} shows the resulting controls and state propagation for basis state $\ket{00}$ obtained in simulation by iLQR. The infidelity is $5.9\cdot 10^{-5}$ with a gate duration of $240\:$ns. The finer grid search includes three runs with infidelities better than $10^{-4}$. None of them have smoother controls or more direct state population probabilities.
			%With the grid search, we were not able to find smoother pulses or more direct state population probabilities.
			%We use the derivative control to ensure a smoother rise and decline in the amplitudes. 
			
			\section{Conclusion}\label{sec:conclusion} % ~0.5 page
			%%% Contribution
			%\elie{basis or native?}  \dirk{IBM uses ECR gate in their native gate set, which is a different matrix than the CR matrix. They use it because they cancel certain terms. Historically, CR gate was the previous gate implementation on fixed-frequency superconducting}
			In this work, we demonstrated that iLQR can be used in quantum optimal control, especially for superconducting gate synthesis. We have focused on the single- and two-qubit operations, X and CR gates, among the basis gates for fixed-frequency transmon qubits. The algorithm optimizes the derivatives indirectly influencing the controls to ensure smoother piecewise constant controls. Additionally, this allows us to constrain the start and end controls closer to zero. In particular, the optimizer finds smooth solutions with high gate fidelities for the $X$ gate $40\:$ns pulses, and the $CR$ gate $240\:$ns pulse controls. The design of the quadratic cost has had more impact on the results than the initial pulse shapes. Hence, our iLQR setup leads to high-fidelity solutions without requiring knowledge about an approximate solution. In contrast, the shape of the \acr{L-BFGS-B} GRAPE solution depends heavily on the input pulse. While \acr{L-BFGS-B} GRAPE finds solutions with even higher fidelities for a wide range of initial pulses, for reasonable initial pulses, the solutions tend to stay close to the initial pulse shape. Hence, we see the potential to use iLQR for solving QOC problems where no analytical solution is known a priori.
			
			%\elie{a couple of words about high frequencies?}
			%Calculating the Fourier transformation of the controls reveals the %bandwidth
			
			%%% Outlook
			%Analytical closed-form solution of exp, resp. dexp, are provided for $\operatorname{SU}(2)$, $\operatorname{SU}(3)$ and $\operatorname{SU}(4)$ in~\cite{kaiser2022}, resp. $\operatorname{SU}(2)$ in~\cite{Foroozandeh2021}. Implementing the analytic solution would reduce the computational time.
			%~\cite{berger2024}
			%In addition, the current adaptation of iLQR determines the difference between the target and final unitary by Euclidean distance.
			%%shorter than calibrated pulses on publicly available IBM Eagle systems, and suffer from $T_1$ and $T_2$ noise effects
			The iLQR gates are still relatively long, so optimizing for shorter pulses under realistic AWG constraints is a crucial next step. This might be made possible by using more sophisticated cost functions, for example, time-dependent running control costs for the controls $R_c$ and their derivatives $R_d$ to enforce different pulse envelopes. Additional work is needed to incorporate the fidelity directly in the cost function, which could further improve the solution's fidelity. Moreover, our proposed method can be applied to different hardware system architectures. One example of another superconducting system is fixed-frequency transmons coupled to a tunable transmon, which naturally allows for iSWAP or CZ gates~\cite{stehlik2021} as their fundamental two-qubit operation. Open-loop numerical methods like the one presented in this work rely on accurate model and control signal descriptions, which are especially difficult to satisfy for superconducting transmon quantum systems. One approach to deal with the model's inaccuracy is to incorporate hardware feedback~\cite{wu2018}. As a next step, we will investigate how iLQR can integrate hardware feedback to obtain pulses specialized to specific quantum systems.
			%As a next step, we will investigate how pulses obtained from iLQR can be improved with hardware feedback.
			%incorporating $T_1$ and $T_2$ noise
			%into the simulation are two interesting future directions.
			%Recalibrating the obtained pulses on real system devices
			%is difficult because they are not determined by a simple, underlying function.
			\section*{Acknowledgment}
			This work was funded by the German Ministry for Economic Affairs and Climate Action (BMWK) and German Aerospace Center in the project QuBER-KI (grant numbers 50RA2207A and 50RA2207B) as well as by the German Federal Ministry of Education and Research (BMBF) and the Association of German Engineers (VDI) in the project Q\textsuperscript{3}-UP! (grant numbers 4030112 and 13N15779). T.A. and T.Y.H. acknowledge financial support from the Knut and Alice Wallenberg Foundation through the Wallenberg Center for Quantum Technology (WACQT).
			
			%\clearpage
			\bibliographystyle{IEEEtran}
			\bibliography{ilqr}

\begin{thebibliography}{10}
\providecommand{\url}[1]{#1}
\csname url@rmstyle\endcsname
\providecommand{\newblock}{\relax}
\providecommand{\bibinfo}[2]{#2}
\providecommand\BIBentrySTDinterwordspacing{\spaceskip=0pt\relax}
\providecommand\BIBentryALTinterwordstretchfactor{4}
\providecommand\BIBentryALTinterwordspacing{\spaceskip=\fontdimen2\font plus
\BIBentryALTinterwordstretchfactor\fontdimen3\font minus
  \fontdimen4\font\relax}
\providecommand\BIBforeignlanguage[2]{{%
\expandafter\ifx\csname l@#1\endcsname\relax
\typeout{** WARNING: IEEEtran.bst: No hyphenation pattern has been}%
\typeout{** loaded for the language `#1'. Using the pattern for}%
\typeout{** the default language instead.}%
\else
\language=\csname l@#1\endcsname
\fi
#2}}

\bibitem{glaser2015}
\BIBentryALTinterwordspacing
S.~J. Glaser, U.~Boscain, T.~Calarco, C.~P. Koch, W.~K{\"o}ckenberger,
  \emph{et~al.}, ``Training schr{\"o}dinger’s cat: Quantum optimal control:
  Strategic report on current status, visions and goals for research in
  europe,'' \emph{The European Physical Journal D}, vol.~69, pp. 1--24, 2015.
  [Online]. Available: \url{https://doi.org/10.1140/epjd/e2015-60464-1}
\BIBentrySTDinterwordspacing

\bibitem{liu2022}
\BIBentryALTinterwordspacing
J.~Liu, M.~Zhang, H.~Chen, L.~Wang, and H.~Yuan, ``Optimal scheme for quantum
  metrology,'' \emph{Advanced Quantum Technologies}, vol.~5, no.~1, p. 2100080,
  2022. [Online]. Available: \url{https://doi.org/10.1002/qute.202100080}
\BIBentrySTDinterwordspacing

\bibitem{rembold2020}
\BIBentryALTinterwordspacing
P.~Rembold, N.~Oshnik, M.~M. Müller, S.~Montangero, T.~Calarco, \emph{et~al.},
  ``Introduction to quantum optimal control for quantum sensing with
  nitrogen-vacancy centers in diamond,'' \emph{AVS Quantum Science}, vol.~2,
  no.~2, p. 024701, 06 2020. [Online]. Available:
  \url{https://doi.org/10.1116/5.0006785}
\BIBentrySTDinterwordspacing

\bibitem{koch2022}
\BIBentryALTinterwordspacing
C.~P. Koch, U.~Boscain, T.~Calarco, G.~Dirr, S.~Filipp, \emph{et~al.},
  ``Quantum optimal control in quantum technologies. strategic report on
  current status, visions and goals for research in europe,'' \emph{EPJ Quantum
  Technology}, vol.~9, no.~1, p.~19, 2022. [Online]. Available:
  \url{https://doi.org/10.1140/epjqt/s40507-022-00138-x}
\BIBentrySTDinterwordspacing

\bibitem{reich2012}
\BIBentryALTinterwordspacing
D.~M. Reich, M.~Ndong, and C.~P. Koch, ``{Monotonically convergent optimization
  in quantum control using Krotov's method},'' \emph{The Journal of Chemical
  Physics}, vol. 136, no.~10, p. 104103, 03 2012. [Online]. Available:
  \url{https://doi.org/10.1063/1.3691827}
\BIBentrySTDinterwordspacing

\bibitem{krotov1993}
\BIBentryALTinterwordspacing
V.~F. Krotov, \emph{Global Methods in Optimal Control Theory}.\hskip 1em plus
  0.5em minus 0.4em\relax Boston, MA: Birkh{\"a}user Boston, 1993, pp. 74--121.
  [Online]. Available: \url{https://doi.org/10.1007/978-1-4612-0349-0_3}
\BIBentrySTDinterwordspacing

\bibitem{shao2022}
\BIBentryALTinterwordspacing
J.~Shao, J.~Combes, J.~Hauser, and M.~M. Nicotra, ``Projection-operator-based
  newton method for the trajectory optimization of closed quantum systems,''
  \emph{Phys. Rev. A}, vol. 105, p. 032605, 03 2022. [Online]. Available:
  \url{https://link.aps.org/doi/10.1103/PhysRevA.105.032605}
\BIBentrySTDinterwordspacing

\bibitem{hauser2002}
\BIBentryALTinterwordspacing
J.~Hauser, ``A projection operator approach to the optimization of trajectory
  functionals,'' \emph{IFAC Proceedings Volumes}, vol.~35, no.~1, pp. 377--382,
  2002, 15th IFAC World Congress. [Online]. Available:
  \url{https://www.sciencedirect.com/science/article/pii/S1474667015387334}
\BIBentrySTDinterwordspacing

\bibitem{trowbridge2023}
\BIBentryALTinterwordspacing
A.~Trowbridge, A.~Bhardwaj, K.~He, D.~I. Schuster, and Z.~Manchester, ``Direct
  collocation for quantum optimal control,'' in \emph{2023 IEEE International
  Conference on Quantum Computing and Engineering (QCE)}.\hskip 1em plus 0.5em
  minus 0.4em\relax Los Alamitos, CA, USA: IEEE Computer Society, 09 2023, pp.
  1278--1285. [Online]. Available:
  \url{https://doi.ieeecomputersociety.org/10.1109/QCE57702.2023.00144}
\BIBentrySTDinterwordspacing

\bibitem{hargraves1987}
\BIBentryALTinterwordspacing
C.~R. Hargraves and S.~W. Paris, ``Direct trajectory optimization using
  nonlinear programming and collocation,'' \emph{AIAA J. Guidance}, vol.~10,
  pp. 338--342, 1987. [Online]. Available:
  \url{https://doi.org/10.2514/3.20223}
\BIBentrySTDinterwordspacing

\bibitem{li2004}
\BIBentryALTinterwordspacing
W.~Li and E.~Todorov, ``Iterative linear quadratic regulator design for
  nonlinear biological movement systems,'' in \emph{First International
  Conference on Informatics in Control, Automation and Robotics}, vol.~2.\hskip
  1em plus 0.5em minus 0.4em\relax SciTePress, 2004, pp. 222--229. [Online].
  Available: \url{https://doi.org/10.5220/0001143902220229}
\BIBentrySTDinterwordspacing

\bibitem{wensing2024}
\BIBentryALTinterwordspacing
P.~M. Wensing, M.~Posa, Y.~Hu, A.~Escande, N.~Mansard, \emph{et~al.},
  ``Optimization-based control for dynamic legged robots,'' \emph{IEEE
  Transactions on Robotics}, vol.~40, pp. 43--63, 2024. [Online]. Available:
  \url{https://doi.org/10.1109/TRO.2023.3324580}
\BIBentrySTDinterwordspacing

\bibitem{motzoi2009}
\BIBentryALTinterwordspacing
F.~Motzoi, J.~M. Gambetta, P.~Rebentrost, and F.~K. Wilhelm, ``Simple pulses
  for elimination of leakage in weakly nonlinear qubits,'' \emph{Phys. Rev.
  Lett.}, vol. 103, p. 110501, 9 2009. [Online]. Available:
  \url{https://link.aps.org/doi/10.1103/PhysRevLett.103.110501}
\BIBentrySTDinterwordspacing

\bibitem{chow2011}
\BIBentryALTinterwordspacing
J.~M. Chow, A.~D. C\'orcoles, J.~M. Gambetta, C.~Rigetti, B.~R. Johnson,
  \emph{et~al.}, ``Simple all-microwave entangling gate for fixed-frequency
  superconducting qubits,'' \emph{Phys. Rev. Lett.}, vol. 107, p. 080502, 8
  2011. [Online]. Available:
  \url{https://link.aps.org/doi/10.1103/PhysRevLett.107.080502}
\BIBentrySTDinterwordspacing

\bibitem{sheldon2016}
\BIBentryALTinterwordspacing
S.~Sheldon, E.~Magesan, J.~M. Chow, and J.~M. Gambetta, ``Procedure for
  systematically tuning up cross-talk in the cross-resonance gate,''
  \emph{Phys. Rev. A}, vol.~93, p. 060302, 6 2016. [Online]. Available:
  \url{https://link.aps.org/doi/10.1103/PhysRevA.93.060302}
\BIBentrySTDinterwordspacing

\bibitem{jurcevic2021}
\BIBentryALTinterwordspacing
P.~Jurcevic, A.~Javadi-Abhari, L.~S. Bishop, I.~Lauer, D.~F. Bogorin,
  \emph{et~al.}, ``Demonstration of quantum volume 64 on a superconducting
  quantum computing system,'' \emph{Quantum Science and Technology}, vol.~6,
  no.~2, p. 025020, 3 2021. [Online]. Available:
  \url{https://dx.doi.org/10.1088/2058-9565/abe519}
\BIBentrySTDinterwordspacing

\bibitem{khaneja2005}
\BIBentryALTinterwordspacing
N.~Khaneja, T.~Reiss, C.~Kehlet, T.~Schulte-Herbr{\"u}ggen, and S.~J. Glaser,
  ``Optimal control of coupled spin dynamics: design of nmr pulse sequences by
  gradient ascent algorithms,'' \emph{Journal of magnetic resonance}, vol. 172,
  no.~2, pp. 296--305, 2005. [Online]. Available:
  \url{https://doi.org/10.1016/j.jmr.2004.11.004}
\BIBentrySTDinterwordspacing

\bibitem{machnes2011}
\BIBentryALTinterwordspacing
S.~Machnes, U.~Sander, S.~J. Glaser, P.~de~Fouqui\`eres, A.~Gruslys,
  \emph{et~al.}, ``Comparing, optimizing, and benchmarking quantum-control
  algorithms in a unifying programming framework,'' \emph{Phys. Rev. A},
  vol.~84, p. 022305, 8 2011. [Online]. Available:
  \url{https://link.aps.org/doi/10.1103/PhysRevA.84.022305}
\BIBentrySTDinterwordspacing

\bibitem{johansson2012}
\BIBentryALTinterwordspacing
J.~Johansson, P.~Nation, and F.~Nori, ``Qutip 2: A python framework for the
  dynamics of open quantum systems,'' \emph{Computer Physics Communications},
  vol. 184, no.~4, pp. 1234--1240, 2013. [Online]. Available:
  \url{https://www.sciencedirect.com/science/article/pii/S0010465512003955}
\BIBentrySTDinterwordspacing

\bibitem{matekole2022}
\BIBentryALTinterwordspacing
E.~S. Matekole, Y.-L.~L. Fang, and M.~Lin, ``Methods and results for quantum
  optimal pulse control on superconducting qubit systems,'' in \emph{2022 IEEE
  International Parallel and Distributed Processing Symposium Workshops
  (IPDPSW)}, 2022, pp. 600--606. [Online]. Available:
  \url{https://doi.org/10.1109/IPDPSW55747.2022.00102}
\BIBentrySTDinterwordspacing

\bibitem{eitan2011}
\BIBentryALTinterwordspacing
R.~Eitan, M.~Mundt, and D.~J. Tannor, ``Optimal control with accelerated
  convergence: Combining the krotov and quasi-newton methods,'' \emph{Phys.
  Rev. A}, vol.~83, p. 053426, 05 2011. [Online]. Available:
  \url{https://link.aps.org/doi/10.1103/PhysRevA.83.053426}
\BIBentrySTDinterwordspacing

\bibitem{tannor1992}
\BIBentryALTinterwordspacing
D.~J. Tannor, V.~Kazakov, and V.~Orlov, \emph{Control of Photochemical
  Branching: Novel Procedures for Finding Optimal Pulses and Global Upper
  Bounds}.\hskip 1em plus 0.5em minus 0.4em\relax Boston, MA: Springer US,
  1992, pp. 347--360. [Online]. Available:
  \url{https://doi.org/10.1007/978-1-4899-2326-4_24}
\BIBentrySTDinterwordspacing

\bibitem{goerz2019}
\BIBentryALTinterwordspacing
M.~H. Goerz, D.~Basilewitsch, F.~Gago-Encinas, M.~G. Krauss, K.~P. Horn,
  \emph{et~al.}, ``Krotov: A python implementation of krotov's method for
  quantum optimal control,'' \emph{SciPost Phys.}, vol.~7, p. 080, 2019.
  [Online]. Available: \url{https://scipost.org/10.21468/SciPostPhys.7.6.080}
\BIBentrySTDinterwordspacing

\bibitem{mayne1966}
\BIBentryALTinterwordspacing
D.~Mayne, ``A second-order gradient method for determining optimal trajectories
  of non-linear discrete-time systems,'' \emph{International Journal of
  Control}, vol.~3, no.~1, pp. 85--95, 1966. [Online]. Available:
  \url{https://doi.org/10.1080/00207176608921369}
\BIBentrySTDinterwordspacing

\bibitem{tassa2014}
Y.~Tassa, N.~Mansard, and E.~Todorov, ``Control-limited differential dynamic
  programming,'' in \emph{2014 IEEE International Conference on Robotics and
  Automation (ICRA)}, 2014, pp. 1168--1175.

\bibitem{majer2007}
\BIBentryALTinterwordspacing
J.~Majer, J.~Chow, J.~Gambetta, J.~Koch, B.~Johnson, \emph{et~al.}, ``Coupling
  superconducting qubits via a cavity bus,'' \emph{Nature}, vol. 449, no. 7161,
  pp. 443--447, 2007. [Online]. Available:
  \url{https://doi.org/10.1038/nature06184}
\BIBentrySTDinterwordspacing

\bibitem{krantz2019}
\BIBentryALTinterwordspacing
P.~Krantz, M.~Kjaergaard, F.~Yan, T.~P. Orlando, S.~Gustavsson, \emph{et~al.},
  ``A quantum engineer's guide to superconducting qubits,'' \emph{Applied
  physics reviews}, vol.~6, no.~2, 2019. [Online]. Available:
  \url{https://doi.org/10.1063/1.5089550}
\BIBentrySTDinterwordspacing

\bibitem{magesan2020}
\BIBentryALTinterwordspacing
E.~Magesan and J.~M. Gambetta, ``Effective hamiltonian models of the
  cross-resonance gate,'' \emph{Phys. Rev. A}, vol. 101, p. 052308, 5 2020.
  [Online]. Available:
  \url{https://link.aps.org/doi/10.1103/PhysRevA.101.052308}
\BIBentrySTDinterwordspacing

\bibitem{alexander2020}
\BIBentryALTinterwordspacing
T.~Alexander, N.~Kanazawa, D.~J. Egger, L.~Capelluto, C.~J. Wood,
  \emph{et~al.}, ``Qiskit pulse: programming quantum computers through the
  cloud with pulses,'' \emph{Quantum Science and Technology}, vol.~5, no.~4, p.
  044006, 2020. [Online]. Available:
  \url{https://doi.org/10.1088/2058-9565/aba404}
\BIBentrySTDinterwordspacing

\bibitem{mcKay2017}
\BIBentryALTinterwordspacing
D.~C. McKay, C.~J. Wood, S.~Sheldon, J.~M. Chow, and J.~M. Gambetta,
  ``Efficient $z$ gates for quantum computing,'' \emph{Phys. Rev. A}, vol.~96,
  p. 022330, 08 2017. [Online]. Available:
  \url{https://link.aps.org/doi/10.1103/PhysRevA.96.022330}
\BIBentrySTDinterwordspacing

\bibitem{kaiser2022}
\BIBentryALTinterwordspacing
N.~Kaiser, ``Solving the matrix exponential function for the lie groups su(3),
  su(4) and sp(2),'' \emph{The European Physical Journal A}, vol.~58, no. 170,
  09 2022. [Online]. Available:
  \url{https://doi.org/10.1140/epja/s10050-022-00816-5}
\BIBentrySTDinterwordspacing

\bibitem{mastalli2022}
\BIBentryALTinterwordspacing
C.~Mastalli, W.~Merkt, J.~Marti-Saumell, H.~Ferrolho, J.~Sol{\`a},
  \emph{et~al.}, ``A feasibility-driven approach to control-limited ddp,''
  \emph{Autonomous Robots}, vol.~46, no.~8, pp. 985--1005, 2022. [Online].
  Available: \url{https://doi.org/10.1007/s10514-022-10061-w}
\BIBentrySTDinterwordspacing

\bibitem{stehlik2021}
\BIBentryALTinterwordspacing
J.~Stehlik, D.~M. Zajac, D.~L. Underwood, T.~Phung, J.~Blair, \emph{et~al.},
  ``Tunable coupling architecture for fixed-frequency transmon superconducting
  qubits,'' \emph{Phys. Rev. Lett.}, vol. 127, p. 080505, 08 2021. [Online].
  Available: \url{https://link.aps.org/doi/10.1103/PhysRevLett.127.080505}
\BIBentrySTDinterwordspacing

\bibitem{wu2018}
\BIBentryALTinterwordspacing
R.-B. Wu, B.~Chu, D.~H. Owens, and H.~Rabitz, ``Data-driven gradient algorithm
  for high-precision quantum control,'' \emph{Phys. Rev. A}, vol.~97, p.
  042122, 04 2018. [Online]. Available:
  \url{https://link.aps.org/doi/10.1103/PhysRevA.97.042122}
\BIBentrySTDinterwordspacing

\end{thebibliography}
			
		\end{document}